\documentclass[aps,prl,twocolumn,superscriptaddress]{revtex4-1}

\usepackage{graphicx}
\usepackage{amssymb}
\usepackage{amsmath}
\usepackage{color}

\begin{document}

\title{Observation of a crossover from nodal to gapped superconductivity in Lu$_x$Zr$_{1-x}$B$_{12}$}

\author{Franziska~K.~K.~Kirschner}
\email{franziska.kirschner@physics.ox.ac.uk}
\affiliation{Department of Physics, University of Oxford, Clarendon Laboratory, Parks Road, Oxford, OX1 3PU, United Kingdom}
\author{Nikolay~E.~Sluchanko}
\affiliation{A.~M.~Prokhorov General Physics Institute of RAS, 38 Vavilov Street, 119991 Moscow, Russia}
\affiliation{Moscow Institute of Physics and Technology, Institutskii per. 9, Dolgoprudnyi, Moscow Region 141700, Russia}
\author{Vladimir~B.~Filipov}
\affiliation{Frantsevich Institute for Problems of Materials Science NAS, 3, Krzhyzhanovsky str., Kiev 03680, Ukraine}
 \author{Francis~L.~Pratt}
\affiliation{ISIS Facility, STFC Rutherford Appleton Laboratory, Chilton, Didcot, Oxfordshire OX11 0QX, United Kingdom}
 \author{Chris~Baines}
\affiliation{Laboratory for Muon Spin Spectroscopy, Paul Scherrer Institute, CH-5232 Villigen, PSI, Switzerland}
\author{Natalya~Yu.~Shitsevalova}
\affiliation{Frantsevich Institute for Problems of Materials Science NAS, 3, Krzhyzhanovsky str., Kiev 03680, Ukraine}
\author{Stephen J.~Blundell}
\email{stephen.blundell@physics.ox.ac.uk}
\affiliation{Department of Physics, University of Oxford, Clarendon Laboratory, Parks Road, Oxford, OX1 3PU, United Kingdom}
\date{\today}

\begin{abstract}

We have determined the superconducting and magnetic properties of four samples of Lu$_x$Zr$_{1-x}$B$_{12}$ ($x=0.04$, $0.07$, $0.17$, and $0.8$) using muon spin rotation ($\mu$SR) and magnetometry measurements. We observed a strong magnetic signal in both the $\mu$SR and magnetometry data in one sample ($x=0.07$), likely caused by the formation of static moments of size $\approx 1\,\mu_{\rm B}$ due to a clustering effect of the Lu$^{3+}$ ions. In all other samples, we find only a small magnetic signal in the $\mu$SR data thought to originate from boron nuclei in the B$_{12}$ cages. The superconductivity is found to evolve with $x$, with a decrease in $x$ resulting in an increase in critical temperature and a decrease of the penetration depth. Most remarkably, we find the formation of nodes in the superconducting gap for $x \leq 0.17$, providing a new example of an $s$-to-$d$-wave crossover in a superconductor.

\end{abstract}

\maketitle

\emph{Introduction ---} Since the discovery of superconductivity in MgB$_2$ \cite{Nagamatsu2001}, there has been a great interest in searching for superconductivity in a wide class of borides.

ZrB$_{12}$ is BCS superconductor with one of the highest critical temperatures ($T_{\rm c}=6$\,K) among the higher borides \cite{Matthias1968}. LuB$_{12}$, another dodecaboride, is also a superconductor, albeit with a much lower $T_{\rm c}=0.48$\,K \cite{Matthias1968} despite having a very similar crystal structure \cite{Kuzma1983, Dudka2017}, electronic density of states \cite{Shein2003, Jager2006, Teyssier2007}, and phonon density of states \cite{Werheit2006, Teyssier2008} to ZrB$_{12}$. It is thought that the high $T_{\rm c}$ in ZrB$_{12}$ originates from the soft vibrations of Zr$^{4+}$ ions in the boron cages, which are responsible for considerable electron-phonon interactions \cite{Lortz2005, Rybina2010, Sluchanko2013}. In LuB$_{12}$, vibrations of Lu$^{3+}$ ions have almost no contribution to the electron-phonon coupling, perhaps due to the `volume filling factor' of the B cages, which tunes the hybridization of the Lu/Zr and B orbitals \cite{Teyssier2008}. Another factor which may contribute to this effect in LuB$_{12}$ is the development of an electron instability due to the formation of dynamic charge stripes \cite{Sluchanko2018}. ZrB$_{12}$ is thought to be a type II superconductor with $\kappa \sim 2^{-0.5}$ \cite{Tsindlekht2004, Wang2005}, and one study has found it to contain additional mixed and intermediate states \cite{Biswas2012}. LuB$_{12}$ appears to display $s$-wave behaviour \cite{Flachbart2005}. There has been considerable debate surrounding the nature of the superconducting gap function in ZrB$_{12}$. It has been suggested that ZrB$_{12}$ is either a single-gap $s$-wave \cite{Daghero2004, Tsindlekht2004}, two-gap $s$-wave \cite{Gasparov2006}, or a $d$-wave \cite{Lortz2005} superconductor, with its Fermi surface composed of one open and two closed sheets \cite{Shein2003, Teyssier2007}. 

Nonmagnetic impurity substitutions impact on superconducting properties in various ways dependent on the pair-breaking mechanism. For example, Anderson's theorem implies that a small number of nonmagnetic impurities can dramatically suppress superconductivity in the case of an anisotropic gap in a $d$-wave superconductor \cite{Anderson1959, Abrikosov1961, Balatsky2006}. Experiments on cuprates reveal that a spinless impurity introduced into the high-temperature superconductor host produces a large and spatially extended alternating magnetic polarization in its vicinity \cite{Alloul2009}. Somewhat analogous behavior was found in Lu$_x$Zr$_{1-x}$B$_{12}$, in which nonmagnetic Lu$^{3+}$ ions are substituted for Zr$^{4+}$ ions. Spin-polarized nanodomains of size $\approx 5\,$\AA, containing moments $\approx 6\mu_{\rm B}$ and nucleated around the Lu$^{3+}$ ions, were found in some crystals \cite{Sluchanko2016}, but not in others \cite{Sluchanko2017}. This is possibly due to details of the distribution of the Lu$^{3+}$ ions in the lattice, or the presence of vacancies (as found in YB$_6$ \cite{Sluchanko2016a}).

In this paper, we report muon spin rotation ($\mu$SR) and magnetometry experiments on four samples from the Lu$_x$Zr$_{1-x}$B$_{12}$ family of superconductors to determine their superconducting and magnetic properties. We focus on samples relatively close to the stoichiometric LuB$_{12}$ and ZrB$_{12}$ end members of this family of compounds, in order to investigate the effect of nonmagnetic substitutions in the low-doping ($\leq 20\%$ substitution, or $x \leq 0.2$ and $x \geq 0.8$) regime. One sample showed evidence of magnetism from the nanodomains postulated in Ref.~\cite{Sluchanko2016}. Remarkably, we find that the increase in $T_{\rm c}$ with decreasing $x$ is accompanied by the formation of nodes in the superconducting gap, similar to those observed in the iron pnictides \cite{Reid2012, Guguchia2015}. We also find that, while the magnetism is sample-dependent, the superconducting properties are robust.

\emph{Experimental Details ---} Four single crystals of Lu$_x$Zr$_{1-x}$B$_{12}$ were investigated in this experiment; these included one `magnetic' sample, which had previously displayed the nanodomain behaviour, and three `nonmagnetic' samples, which displayed no such phenomena. The `magnetic' sample, $x=0.07$, was identical to that used in Ref.~\cite{Sluchanko2016}. The remaining three samples, $x=0.04$, $0.17$, and $0.8$, were `nonmagnetic' and identical to those studied in Ref.~\cite{Sluchanko2017}. Details of the crystal growth are described in Refs.~\cite{Sluchanko2016} and \cite{Sluchanko2017}. 

$\mu$SR experiments \cite{Blundell1999, Yaouanc2011} were performed using a dilution refrigerator and $^3$He sorption cryostat mounted on the MuSR spectrometer at the ISIS pulsed muon facility (Rutherford Appleton Laboratory, UK) \cite{King2013}. Further experiments were carried out using the low temperature facility (LTF) and general purpose spectrometer (GPS) at the Swiss Muon Source. Transverse-field (TF) measurements were made to identify the superconducting ground state and its evolution with $x$. Zero-field (ZF) and longitudinal-field (LF) measurements were carried out in order to test for magnetic phases in the sample. All of the $\mu$SR data were analyzed using WiMDA \cite{Pratt2000}. Magnetometry measurements were carried out on a Quantum Design SQUID magnetometer.

\begin{figure}[t]
	\includegraphics[width=.45\textwidth]{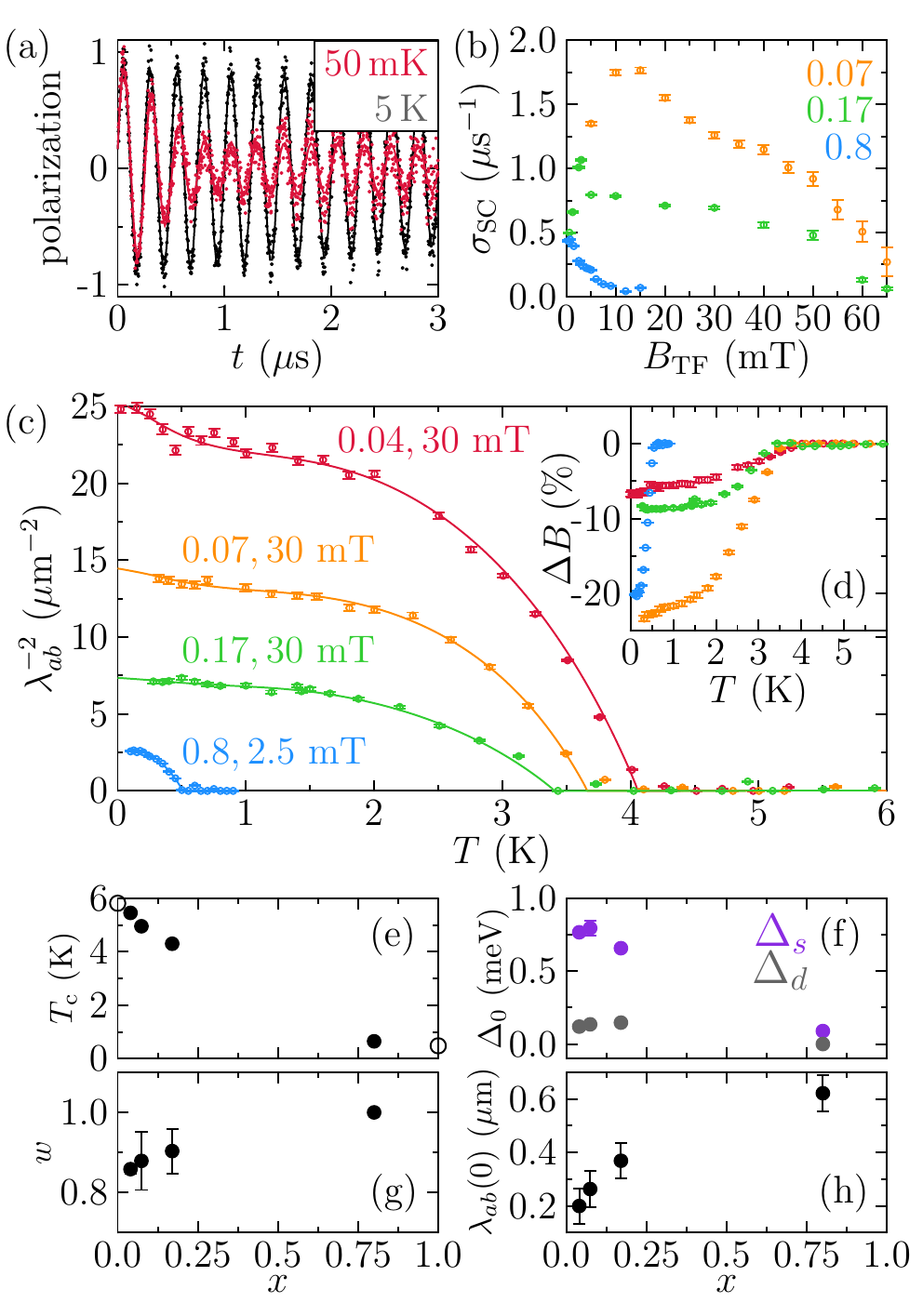}
	\caption{(a) Sample TF-$\mu$SR spectra above and below $T_{\rm c}$ for $x=0.04$. Fits as in Eq.~\ref{TFfit} are also plotted. (b) Field dependence of the relaxation due to superconductivity for $x=0.07$ (orange), $x=0.17$ (green), and $x=0.8$ (blue). (c) Temperature dependence of the inverse square penetration depth, with $s+d$-wave fits for $x=0.04,0.07,$ and $x=0.17$, and an $s$-wave fit for $x=0.8$. (d) $T$ dependence of the field shift due to superconductivity. The evolution of $T_{\rm c}$ (the zero-field value, extrapolated from magnetometry data), the superconducting gaps, the gap ratio, and the penetration depth with $x$ are given in (e), (f), (g) and (h) respectively. The unfilled black points in (e) are the $T_{\rm c}$ values for ZrB$_{12}$ and LuB$_{12}$ from Ref.~\cite{Matthias1968}.}
	\label{TFfig}
\end{figure}

\emph{Superconductivity ---} TF-$\mu$SR measurements were performed on all of the samples to determine their superconducting properties. $x=0.04$, $0.07$, and $0.17$ were measured above and below $T_{\rm c}$ in transverse applied fields $B_{\rm TF}$ of 30\,mT and $x=0.8$ was measured in $B_{\rm TF}=2.5$\,mT [fields were chosen in order to lie below $B_{\rm c2}$ for each compound; see the Supplemental Information \footnote{See Supplemental Material at [URL will be inserted by publisher] for further discussion of the nodal and gapped fitting, and additional plots of the superconducting parameters.} for $B_{\rm c2}$ values]. A sample spectrum for $x=0.04$ is shown in Fig.~\ref{TFfig}(a). Transverse field sweeps were also made to determine the field dependence of the internal field distribution of the superconducting state. 

The data were fitted with
\begin{equation} \label{TFfit}
A = A_{\rm SC} \cos(\gamma_{\mu} B_{\rm SC} t + \phi) e^{-\frac{\sigma^2 t}{2}} + A_{\rm TF} \cos(\gamma_{\mu} B_{\rm TF} t + \phi) e^{-\frac{\sigma_{\rm TF}^2 t}{2}},
\end{equation}
where $\gamma_{\mu} = 2\pi \times 135.5\,\rm{MHzT}^{-1}$ is the gyromagnetic ratio of the muon and $\phi$ is related to the detector geometry [the data were divided among eight groups of detectors with $\phi$ fitted for each group]. The first term corresponds to muons that are not in the superconductor, and precess in the external field. These muons experience a small Gaussian relaxation. The second term corresponds to muons in the superconductor, which experience an approximately Gaussian relaxation arising from the field distribution of the vortex lattice. These muons also experience a small, temperature independent relaxation from nuclear moments, giving a total relaxation $\sigma (T) = \sqrt{\sigma_{\rm SC}^2(T) + \sigma_{\rm nucl}^2}$. 

The transverse field dependences of $\sigma_{\rm SC}$ for $x~=~0.07$, $0.17$, and $0.8$ are shown in Fig.~\ref{TFfig}(b). There is a muon quadrupolar resonance due to the boron nuclei at $\approx~4$\,mT, which may affect the low-field dependence (this will be further discussed in the next section). Apart from $x=0.8$, the upper critical fields $B_{\rm c2}$ of all the samples are larger than the fields resolvable by the experiment; these have instead been calculated from magnetometry and heat capacity measurements in Refs.~\cite{Sluchanko2016} and \cite{Sluchanko2017}, and are shown in \footnotemark[\value{footnote}].

Assuming all of the samples are type~II superconductors with an isotropic hexagonal Abrikosov vortex lattice in the a-b plane that can be described by Ginzburg-Landau theory, the in-plane penetration depth $\lambda_{ab}$ can be extracted from the relaxation due to superconductivity using \cite{Brandt2003} $\sigma_{\rm SC} = 0.0609 \gamma_{\mu}\phi_0 \lambda_{ab}^{-2}(T)$, where $\phi_0 = 2.069 \times 10^{-15}$\,Wb is the magnetic flux quantum.  This formula is an approximation which holds for $0.13/\kappa^2 \ll B_{\rm TF}/B_{\rm c2} \ll 1$ where $\kappa$ is the Ginzburg-Landau parameter \footnotemark[\value{footnote}]. By comparing this approximation to the analytical relationship between $\sigma_{\rm SC}$ and the applied field in Ref.~\cite{Brandt2003}, we find it correctly describes this relationship to within $10\%$ for the $x=0.04$ and $0.07$ samples, and to within $20\%$ for $x=0.17$ and $0.8$. The temperature dependence of $\lambda_{ab}^{-2}$ is shown in Fig.~\ref{TFfig}(c), and the corresponding field shifts due to the vortex lattice are given in Fig.~\ref{TFfig}(d).

To determine the nature of the superconducting gaps in the samples, we fitted the data with single-gap BCS $s$-wave and $d$-wave models, as well as two-gap $s+s$- and $s+d$-wave models. The BCS model of the normalized superfluid density of a superconductor is given by \cite{Chandrasekhar1993}:
\begin{equation}
\tilde{n}_{\rm s}(T) = \frac{\lambda_{ab}^{-2}(T)}{\lambda_{ab}^{-2}(0)} = 1 + \frac{1}{\pi} \int^{2\pi}_{0} \int^{\infty}_{\Delta(\phi,T)} \frac{\partial f}{\partial E} \frac {E {\rm d}E \rm{d}\phi}{\sqrt{E^2 - \Delta^2 (\phi,T)}},
\end{equation}
where $\Delta(\phi,T)$ is the superconducting gap function, and $f=\left[1+\exp \left(E/k_{\rm B}T \right)\right]^{-1}$ is the Fermi function. The gap function can be approximated as $\Delta(\phi,T) = \Delta(\phi) \tanh \left(1.82 \left[ 1.018 \left(T_{\rm c}/T - 1 \right) \right]^{0.51} \right)$. The angular gap function $\Delta(\phi)=\Delta_0$ for $s$-wave superconductors and $\Delta(\phi) = \Delta_0 \cos (2\phi)$ for $d$-wave (nodal) superconductors. For multigap models, the total $\tilde{n}_{\rm s}(T)$ is given by a weighted sum of the superfluid densities for the individual gaps: $\tilde{n}_{\rm s}(T) = w\tilde{n}_{\rm s}^{\rm gap~1}(T) + (1-w)\tilde{n}_{\rm s}^{\rm gap~2}(T)$. In the case of the $s+d$-wave fits, the first gap is $s$-wave, whereas the second gap is $d$-wave.

We find that $x=0.04$, $0.07$, and $0.17$ are $s+d$-wave superconductors, whereas $x=0.8$ is purely $s$-wave. A key indicator of nodal superconductivity is the linear dependence of $\lambda_{ab}^{-2}$ at low $T$, which is observed in $x=0.04$, $0.07$, and $0.17$. $x=0.8$, on the other hand, shows a low temperature plateau in $\lambda_{ab}^{-2}$, corresponding to a fully gapped superconductor where low-energy excitations are strongly suppressed \cite{Graf1995}. This is consistent with the $s$-wave superconductivity observed in LuB$_{12}$ \cite{Flachbart2005}. Further discussion of this fitting can be found in the Supplemental Information \footnotemark[\value{footnote}]. Our observations give clear evidence for the opening of a superconducting gap as Lu$^{3+}$ concentration increases, leading to the rapid suppression of $T_{\rm c}$, shown in Fig.~\ref{TFfig}(e). The fitted values of the $s$- and $d$-wave superconducting gaps, $\Delta_s$ and $\Delta_d$ respectively, are shown in Fig,~\ref{TFfig}(f); we find that along with the complete suppression of the nodal gap (indicated by $w \to 1$ in Fig.~\ref{TFfig}(g)), the $s$-wave gap also decreases significantly in size with increasing $x$. As $x$ becomes larger, the superfluid density ($n_{\rm s} \propto \lambda_{ab}^{-2}(0)$) is suppressed, leading to a longer penetration depth as shown in Fig.~\ref{TFfig}(h). This agrees well with the two peaks observed in the field dependence of $\sigma_{\rm SC}$ for $x=0.07$ (see Fig.~\ref{TFfig}(b)). The peak at $B_{\rm TF} \approx 12$\,mT may be a reflection of the field dependence of the weak $d$-wave term on top of the $s$-wave term, as $d$-wave contribution is more easily suppressed by an external field.

Zr$^{4+}$ has an extra $d$-orbital vacancy compared to Lu$^{3+}$; it is likely that this results in increased $d$-wave orbital pairing with the B cages, and therefore nodal superconductivity dominates. The observed gap evolution in the Lu$_x$Zr$_{1-x}$B$_{12}$ is similar to that seen in Fe-based superconductors. The Ba$_{1-x}$K$_x$Fe$_2$As$_2$ family of materials shows fully gapped behaviour for $x=0.4$, with line nodes appearing at $x=1$ \cite{Reid2012}. A similar transition is achieved in Ba$_{0.65}$Rb$_{0.35}$Fe$_2$As$_2$ using hydrostatic pressure, which promotes a nodal gap \cite{Guguchia2015}. These results indicate that the Lu$_x$Zr$_{1-x}$B$_{12}$ family  may share more similarity with iron pnictide superconductors than with the cuprates. We remark that the superconducting properties of the Lu$_x$Zr$_{1-x}$B$_{12}$ family appear to be much more sensitive to doping than the 122 iron-based superconductors.

\emph{Magnetism ---} To address the apparent discrepancy between the differing magnetic behaviours of this family of compounds observed in Refs. \cite{Sluchanko2016} and \cite{Sluchanko2017}, ZF- and LF-$\mu$SR experiments were carried out.

\begin{figure}[t]
	\includegraphics[width=.45\textwidth]{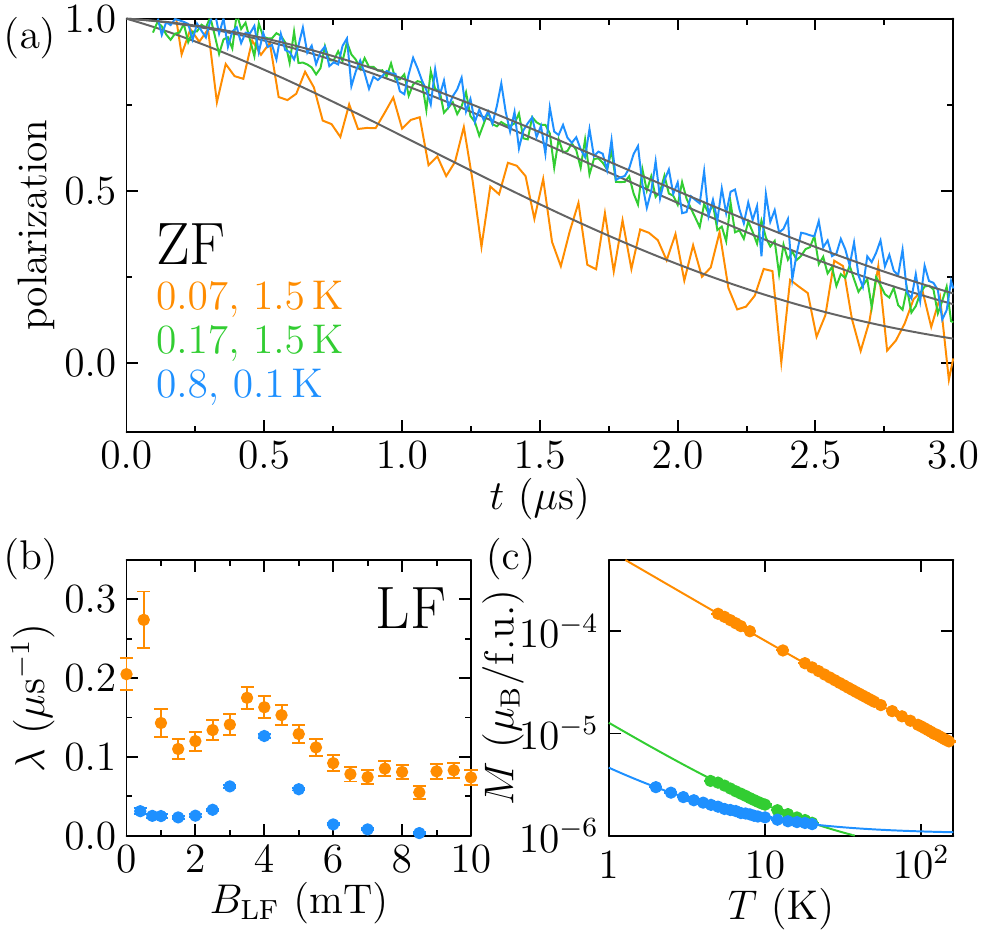}
	\caption{(a) Sample ZF-$\mu$SR spectra in the superconducting state. (b) The field dependence of the relaxation for one `magnetic' sample ($x=0.07$, orange) and one `nonmagnetic' sample ($x=0.8$, blue). (c) The temperature dependence of the magnetization per formula unit.}
	\label{ZFfig}
\end{figure}

ZF spectra below $T_{\rm c}$ for one `magnetic' sample ($x=0.07$) and two `nonmagnetic' samples ($x=0.17$ and $0.8$) are shown in Fig.~\ref{ZFfig}(a). For all samples the ZF spectrum did not change as $T$ was increased above $T_{\rm c}$, meaning that any magnetism present in the samples does not compete with the superconducting phase. No oscillations were seen in the forward-backward asymmetry spectra, and there were also no discontinuous jumps in either the initial or the baseline asymmetry, which suggests there is no long-range order inside the sample. The data were fitted with the zero-field Kubo-Toyabe function \cite{Kubo1967} with an additional Lorentzian relaxation:
\begin{equation}
A = A_0 \left( \frac{1}{3} + \frac{2}{3} \left( 1 - \Delta_{\rm ZF}^2 t^2 \right) e^{-\Delta_{\rm ZF}^2 t^2 / 2} \right) e^{- \lambda_{\rm ZF} t} + A_{\rm b},
\end{equation}
where $A_0$ and $A_{\rm base}$ are the relaxing and baseline amplitudes. The Kubo-Toyabe function is often used to describe a system of static spins characterised by a width $\Delta_{\rm ZF}/\gamma_{\mu}$.

$\Delta_{\rm ZF}$ was found to be constant ($0.40(3)\,\mu\rm{s}^{-1}$) across all of the samples. The additional Lorentzian relaxation was equal and very small for the two nonmagnetic samples ($\approx~0.05\,\mu\rm{s}^{-1}$), and significantly larger for the magnetic sample ($\approx~0.2\,\mu\rm{s}^{-1}$).

To further probe the nature of the magnetism, LF-$\mu$SR was carried out on one magnetic and one non-magnetic sample: $x=0.07$ and $x=0.8$ respectively. The data were fitted with a longitudinal-field Kubo-Toyabe function \cite{Hayano1979} with an additional Lorentzian relaxation $e^{\lambda t}$; the field dependence of $\lambda$ is plotted in Fig.~\ref{ZFfig}(b). Both samples have a peak in their relaxation at $B \approx 4\,\rm{mT}$, corresponding to the muon experiencing a quadrupolar resonance with the $^{11}$B nucleus \cite{Marict1991,Cox1992}. The additional relaxation seen in the ZF spectrum of the magnetic sample is quenched above $\approx 1\,\rm{mT}$, indicating an internal field on the order of 0.1\,mT arising from static magnetic moments [an applied longitudinal field $B_{\rm LF} > 10B_{\rm internal}$ rapidly quenches relaxation arising from static moments \cite{Hayano1979, Yaouanc2011}].

The observed differences in magnetism are supported by bulk magnetometry measurements of $x=0.07$, $0.17$, and $0.8$, as shown in Fig.~\ref{ZFfig}(c). The data were fitted to a Curie-Weiss function with an additional constant background magnetisation:
\begin{equation}
M = M_0 + \mu^2 H / 3k_{\rm B} T,
\end{equation}
where $\mu$ is the effective magnetic moment per formula unit, and $H=20\,\rm{Oe}$ is the applied external field. We find that the moment per formula unit in the magnetic sample ($\mu_{0.07} \approx 1.18 \mu_{\rm B}$) is significantly higher than for the nonmagnetic samples ($\mu_{0.17} \approx 0.2 \mu_{\rm B}$, $\mu_{0.8}~\approx~0.1 \mu_{\rm B}$).

From these data, we postulate that the ZF signal observed in the nonmagnetic samples, which is insensitive to Lu$^{3+}$ concentration, is dominated by B nuclei in the cages, similar to that observed in LuB$_{12}$ \cite{Kalvius2003}. In addition to this nuclear signal, the magnetic sample also contains static, disordered moments, which may potentially be associated with clusters of Lu$^{3+}$, similar to the phenomena discussed in Refs. \cite{Sluchanko2016} and \cite{Sluchanko2016a}. As the apparent clustering effect is only present in one sample, we conclude that the differences in magnetism we observe are due to the detailed distribution of cations that is established during sample preparation, rather than an intrinsic effect linked to $x$.

\emph{Conclusion ---} In summary, we have carried out $\mu$SR and magnetometry measurements to probe the nature of the magnetism and superconductivity in the Lu$_x$Zr$_{1-x}$B$_{12}$ family of superconductors. ZF-$\mu$SR, LF-$\mu$SR, and magnetometry measurements reveal a strong magnetic signal in one sample, which was thought to contain magnetic nanodomains, compared to those samples with no domains. We attribute this sample dependence to a clustering effect in which the distribution of Lu$^{3+}$ may affect the formation of static moments. We find that the moments associated with this effect are of order $1\,\mu_{\rm B}$ per formula unit. Scanning tunnelling microscopy or scanning electron microscopy may provide further insights into this effect. TF-$\mu$SR measurements revealed that the superconductivity was robust to these variations in magnetism. We find that the increase in $T_{\rm c}$ associated with the decrease of $x$ is also accompanied by a decrease in penetration depth and the formation of nodes in the superconducting gap function. This unusual transition is similar to that observed in iron pnictide superconductors, and further study of this family of compounds may provide additional insights into the mechanisms behind nodal-to-gapped crossovers in high temperature superconductors.

\emph{Acknowledgements ---} We thank C.~Topping and F.~Lang for experimental assistance and M.~Bristow and R.~Fernandes for insightful discussions. F.K.K.K. thanks Lincoln College, Oxford, for a doctoral studentship. This study was supported by the program ``Fundamental problems of high-temperature superconductivity'' of the Presidium of the RAS. Part of this work was performed at the Science and Technology Facilities Council (STFC) ISIS Facility, Rutherford Appleton Laboratory, and part at S$\mu$S, the Swiss Muon Source (PSI, Switzerland).

\bibliographystyle{apsrev4-1}
\bibliography{bib1}

\end{document}